\documentclass{article}
\usepackage{graphicx}
\title{\bf Phantom Inflation and the "Big Trip"}
\author{Pedro F. Gonz\'{a}lez-D\'{\i}az\footnote{E-mail:
p.gonzalezdiaz@imaff.cfmac.csic.es}, and Jos\'{e} A. Jim\'{e}nez-Madrid.\\
Colina de los Chopos, Instituto de Matem\'{a}ticas y F\'{\i}sica
Fundamental\\ Consejo Superior de Investigaciones Cient\'{\i}ficas\\
Serrano 121, 28006 Madrid, SPAIN\\ }
\date{June 4, 2004}
\begin{document}
\maketitle \large \setlength{\baselineskip}{0.9cm}

\begin{center}
{\bf Abstract}
\end{center}
Primordial inflation is regarded to be driven by a phantom field
which is here implemented as a scalar field satisfying an equation
of state $p=\omega\rho$, with $\omega<-1$. Being even aggravated
by the weird properties of phantom energy, this will pose a
serious problem with the exit from the inflationary phase. We
argue however in favor of the speculation that a smooth exit from
the phantom inflationary phase can still be tentatively recovered
by considering a multiverse scenario where the primordial phantom
universe would travel in time toward a future universe filled with
usual radiation, before reaching the big rip. We call this
transition the "big trip" and assume it to take place with the
help of some form of anthropic principle which chooses our current
universe as being the final destination of the time transition.

\vspace{.3cm}


\vspace{.3cm}


\vspace{.5cm}

\pagebreak

Phantom energy might be currently dominating the universe. This is
a possibility which is by no means excluded by present constraints
on the universal equation of state [1]. If phantom energy
currently dominated over all other cosmic components in the
universe, then we were starting a period of super-accelerated
inflation which would even be more accelerated than the
accelerating process that corresponded to the existence of a
positive cosmological constant. One could thus be tempted to look
at the primordial inflation as being originated by a phantom
field. In fact, Piao and Zhang have already considered [2] an
early-universe scenario where the inflationary mechanism is driven
by phantom energy. The main difficulty with that scenario is that
it is difficult to figure out in it how the universe can smoothly
exit from the inflationary period, as an inflating phantom
universe seems to inexorably ends up in a catastrophic big rip
singularity. Now, however, we seem to have a tool which might help
to solve that difficulty. It is based on the rather astonishing
effects that accretion of phantom energy can induce in the
evolution of Lorentzian wormholes [3].

Since phantom energy violates dominant energy condition, wormholes
can naturally occur [4] in a universe dominated by phantom energy
[5]. It can in fact be thought that if Planck-sized wormholes are
quantum-mechanically stable and were also allowed to exist in the
primordial spacetime foam of a phantom dominated early universe,
then before occurrence of the big rip [6], the throat of the
wormholes will rapidly increase and become larger than the size of
the inflating universe itself, to blow up before the big rip [3].
The moment at which all the originally Planck-sized wormholes of
the spacetime foam become simultaneously infinite could thus be
dubbed as the "Big Trip" (Fig. 1), as a sufficiently inflated
early universe would then be inside the throat of the wormholes
and could thus be itself converted into a "time traveller" which
may be instantly transferred either to its past or to its future.
Conditions for the Big Trip does not just occur at the time when
the size of the wormhole throat blows up, but they extend
backwards, down to the time at which the wormhole throat size just
overcomes the phantom universe size, all the way along the time
interval $\Delta T_{nonchronal}$ shown in Fig. 1.

In what follows we shall implement the above picture by using a
semi-quantitative scenario where we first describe how a set of
parallel quantum universes (which would include both the phantom
universe and a universe filled with usual radiation) can be
quantum-mechanically derived, and then discuss the way through
which a time travel from the phantom universe to the Friedmann
universe with usual radiation may lead to a smooth exit from
inflation taking place in the former universe. Let us thus
consider the spacetime manifold $M$ for a little flat FRW universe
with metric
\begin{equation}
ds^2=-N^2 dt^2 +a(t)d\Omega_3^2 ,
\end{equation}
where $N$ is the lapse function resulting from foliating the
manifold $M$, $d\Omega_3^2$ is the metric on the unit
three-sphere, and $a(t)$ is the scale factor. Such a universe will
be filled with dark energy and equipped with a general equation of
state $p=\omega\rho$, where $\omega$ is a parameter which is here
allowed to be time-dependent. We shall then formulate the
quantum-mechanical description of such a universe starting with
its action integral. For this we shall take the most general
Hilbert-Einstein action where, besides considering the scale
factor $a(t)$ and the scalar field $\phi(t)$ to be time-dependent,
and hence appearing in the action as a given combination of $a$,
$\phi$, $\dot{a}$ and $\dot{\phi}$ (with $\dot{}=d/dt$), we will
generally regard the state equation parameter $\omega$ to be
initially time-dependent, too, even though we restrict ourselves
to the case where we always have $\ddot{\omega}=0$.

Differentiating then the Friedmann equation (i.e. Eq. (14) below
which will be in principle defined for a flat FRW universe with
constant $\omega$) with respect to time one can see that, if the
$\omega$ in that equation is nevertheless regarded to be
time-dependent then the scalar curvature can be generalized to a
new expression $\hat{R}$ which thereby is suggested to be given by
\[\hat{R}=R-\frac{\dot{a}\dot{\omega}\ln a^9}{a}  ,\]
where $R$ is the conventional Ricci curvature scalar. In obtaining
this expression we have first assumed $\rho=\rho_0
a^{-3(1+\omega)}$, letting then $\omega\equiv\omega(t)$ in it.
Thus, since the scalar curvature for a flat universe is given by
$6(\ddot{a}+\dot{a}^2/a^2)$, it can be checked that
differentiation of Eq. (14) directly produces the extra term
appearing in the expression for $\hat{R}$. To this generalized
Ricci curvature scalar one should then associate a correspondingly
generalized extrinsic curvature $\hat{K}$. In such a case the
action integral of the manifold $M$ with boundary $\partial M$ can
be written as
\begin{eqnarray}
&&S=\int_M d^4 x\sqrt{-g}\left(\frac{\hat{R}}{16\pi
G}+L_{\phi}\right) -\frac{1}{8\pi G}\int_{\partial M}d^3
x\sqrt{-h}Tr\hat{K}\nonumber\\ &&=\int_{M} d^4
x\sqrt{-g}\;[\frac{1}{16\pi G}\left(R-\frac{\dot{a}\dot{\omega}
\ln a^9}{aN^2}\right)
+\frac{1}{2}(\partial_{\mu}\phi)^2-V(\phi)]\nonumber\\
&&-\frac{1}{8\pi G}\int_{\partial M} d^3 x\sqrt{-h}\; \left(Tr
K-\frac{\dot{\omega}\ln a^{3/2}}{N}\right) ,
\end{eqnarray}
in which $K$ is the conventional expression for the extrinsic
curvature, and $g$ and $h$ denote, respectively, the determinants
of the general four-metric $g_{ij}$ on $M$ and three-metric
$h_{ij}$ on the given hypersurface at the boundary $\partial M$,
characterized by a lapse $N$ and a shift $N_i$ functions. We note
that (i) the above action integral becomes the conventional
Hilbert-Einstein action for $\omega$=Const., (ii) in the surface
integral we have also added a new non-conventional extra term
depending on $\dot{\omega}$ which would represent a transition
between different universes each with a fixed equation of state,
and together with $TrK$ would make the trace of the generalized
extrinsic curvature $Tr\hat{K}$, and finally that (iii) the
momenta conjugate to $a(t)$ and $\omega(t)$ are not separable from
each other even when we have assumed $\ddot{\omega}=0$. In the
case that we consider that, in principle, no particular constant
value for $\omega$ is specified, from this action integral the
Hamiltonian constraint can be derived by taking $\delta S/\delta
N$ and this can then be converted into a Wheeler-DeWitt wave
equation by applying a suitable correspondence principle to the
momenta conjugated to both the scale factor $a(t)$ and the state
equation parameter $\omega(t)$. We do not include here a momentum
conjugate to the scalar field $\phi$ because, if $\omega$ is also
promoted to a dynamic variable and $p=\omega\rho$, then
$\dot{\phi}^2$ and $V(\phi)$ can always be expressed as a single
function of $\omega$ and the energy density, that is in terms of
the two dynamic variables $\omega$ and $a$ (see e.g. Eqs. (12) and
(13) given below). Using a most covenient Euclidean manifold where
$t\rightarrow i\tau$ for our flat geometry, the final form of the
Euclidean action can be reduced to
\begin{equation}
I=-\int Nd\tau \left(-\frac{a\dot{a}^2}{N^2}+
\frac{1}{2}\frac{\dot{\omega}\dot{a}a^2}{N^2} +\ell_p^2\omega a^3
\rho(a,\omega)\right) ,\nonumber
\end{equation}
in which $\ell_p =\sqrt{8\pi G/3}$ is the Planck length, and
$\rho(a,\omega)(=\rho_0 a^{-3(1+\omega)}$ in the explicit simple
model considered below) is the dark energy density. In the gauge
where $N=1$ we have then for the Hamiltonian constraint
\begin{equation}
H=\frac{\delta I}{\delta N}-(1+\omega)\ell_p^2 a^3\rho =
a\dot{a}^2 -\frac{1}{2}\dot{a}\dot{\omega}a^2 -\ell_p^2 a^3
\rho(a,\omega) =0 ,\nonumber
\end{equation}
where the term $(1+\omega)\ell_p^2 a^3\rho$ should be added to
correct the effect of replacing the Lagrangian of the field
$L_{\phi}$ for pressure $p=\omega\rho$ initially in the action
$S$. Had we kept $L_{\phi}=\frac{1}{2}\dot{\phi}^2-V(\phi)$ in the
action $S$, then we had obtained Eq. (4) again by simply applying
just the operator $H=\delta I/\delta N$. It is worth noticing that
in the case $\dot{\omega}=0$, Eq. (4) reduced to the customary
Friedmann equation for flat geometry (i.e. Eq. (14) below), and
that if $\dot{\omega}\neq 0$, even though Eq. (4) looks formally
different of Eq. (14), they should ultimately turn out to be fully
equivalent to each other once the fact that the energy density is
given by $\rho=\rho_0 a^{-3}\exp\left(-3\int
\omega(t)\dot{a}dt/a\right)$ if $\omega\neq 0$, and by
$\rho=\rho_0 a^{-3(1+\omega)}$ if $\dot{\omega}=0$, is taken into
account. These expressions for $\rho(a)$ come from direct
integration of the conservation law for cosmic energy,
$\dot{\rho}=-3(p+\rho)\dot{a}/a=-3(1+\omega)\rho\dot{a}/a$, in the
case that either $\omega$ is taken to be time-dependent even
before integrating, or $\omega$=const., respectively. For the
momenta conjugate to $a$ and $\omega$ we have moreover
\begin{equation}
\pi_a =i\left(\frac{1}{2}\dot{\omega}a^2-2a\dot{a}\right) ,\;\;
\pi_{\omega}=\frac{i}{2}\dot{a}a^2 .\nonumber
\end{equation}
In terms of the momenta $\pi_a$ and $\pi_{\omega}$, the
Hamiltonian constraint can be cast in the form
\begin{equation}
H= \pi_{\omega}^2 -\frac{a}{2}\pi_a\pi_{\omega}
-\frac{\ell_p^2}{4}a^6 \rho(a,\omega) =0.\nonumber
\end{equation}
As it was anticipated before, this Hamiltonian is not separable in
the two considered components of the momentum space.

The necessary quantum description requires a correspondence
principle which in the present case leads to introducing the
following quantum operators
\begin{equation}
\pi_a\rightarrow\hat{\pi}_a =-i\ell_p^2 \frac{\partial}{\partial
a} ,\;\;\; \pi_{\omega}\rightarrow\hat{\pi}_{\omega} =-i\ell_p^2
\frac{\partial}{\partial\omega} ,\nonumber
\end{equation}
which, when brought into the Hamiltonian constraint, allow us to
obtain the following Wheeler-DeWitt equation
\begin{equation}
\left[\frac{\partial^2}{\partial\omega^2}
-\frac{1}{2}a\frac{\partial^2}{\partial\omega\partial a}
+\frac{1}{4}\ell_p^{-2} a^6 \rho(a,\omega)\right]\Psi(a,\omega)
=0, \nonumber
\end{equation}
where $\Psi(a,\omega)$ is an in principle nonseparable wave
functional for the original universe.

Solving this equation is very difficult even for the simplest
conceivable initial boundary conditions. There could be, moreover,
a problem with the operator ordering, which would not be alien to
the kind of quantum cosmological framework we are using. However,
if we tentatively assume for a moment that initially the wave
functional can be written as a separable product of the form
$\Psi(a,\omega)=\exp(-a/\ell_p)\Psi(\omega)$, with which the
Wheeler-DeWitt equation would reduce to
\begin{equation}
\left[\frac{\partial^2}{\partial\omega^2}
+\frac{a}{2\ell_p}\frac{\partial}{\partial\omega}
+\frac{1}{4}\ell_p^{-2} a^6 \rho(a,\omega)\right]\Psi(\omega) =0,
\nonumber
\end{equation}
then, if conservation of the total dark energy $E_T=a^3\rho$ is
also assumed, then this differential equation would describe an
oscillator with a damping force with coefficient
$\lambda=a/(2\ell_p)$ for the squared frequency $\nu^2=a^3
E_T/(4\ell_p^2)$, which admits the general solutions

\noindent {\bf 1.} $E_T <1/(4a)$ ({\it Overdamped regime})
\[\Psi(\omega)=e^{-\frac{a\omega}{4\ell_p}}\left(C_1 e^{\frac{\xi_1
a\omega}{4\ell_p}} +C_2 e^{-\frac{\xi_1 a\omega}{4\ell_p}}\right)
\]

\noindent {\bf 2.} $E_T =1/(4a)$ ({\it Critically damped regime})
\[\Psi(\omega)=e^{-\frac{a\omega}{4\ell_p}}\left(C_1 +C_2
\omega\right) \]

\noindent {\bf 3.} $E_T >1/(4a)$ ({\it Underdamped regime})
\begin{equation}
\Psi(\omega)=e^{-\frac{a\omega}{4\ell_p}}\left[C_1
\cos\left(\frac{\xi_2 a\omega}{4\ell_p}\right) +C_2
\sin\left(-\frac{\xi_2 a\omega}{4\ell_p}\right)\right] ,
\end{equation}
where $C_1$ and $C_2$ are arbitrary constants, $\xi_1=\sqrt{1- 4
E_T a}$ and $\xi_1=\sqrt{4 E_T a -1}$.\footnote{Had we chosen for
$\Psi(a)$ a square integrable function of the form
$\Psi(a)=\exp(-a^2/2\ell_p^2)$ satisfying the no-boundary initial
condition [7] we had obtained the same solutions $\Psi(\omega)$
given by Eqs. (10), but replacing in these all ratios $a/\ell_p$
for $a^2/\ell_p^2$, so admitting a similar interpretation.}

The condition $\Psi=\exp(-a/\ell_p)\Psi(\omega)$, or
$\Psi=\exp(-a^2/2\ell_p^2)\Psi(\omega)$, is nevertheless too much
restrictive and therefore one should then have to solve the full
Wheeler-DeWitt equation with the operators
$i\partial/\partial\omega$ and $i\partial/\partial a$ that
correspond to a classical momenta $\pi_{\omega}$ and $\pi_{a}$ for
suitable initial conditions (including e.g. putting upper and
lower limits to the allowed values of $\omega$). Once one have got
a suitable wave functional $\Psi(a,\omega)$, one should Fourier or
Laplace (depending on whether we work in the Lorentzian or
Euclidean formalism) transform it into the corresponding wave
functional in the $K$-representation $\Phi(\dot{a},\dot{\omega})$
[i.e. extending also to $\omega$-space the $K$-representation in
$a$-space [7] which, in the Euclidean manifold, would be obtained
by path integrating $\Psi(a,\omega)\times
exp(aK+\omega\dot{\omega}\ln a^{3/2})$ over $a$ and $\omega$] to
obtain a set of (presumably discrete) states for the little
universe in terms of allowed quantum eigenvalues of parameter
$\omega$. Thus, if the parameter $\omega$ of the equation of state
is not fixed, one can always rearrange things so that it is this
parameter $\omega$ which can be quantum-mechanically described and
take on a set positive and/or negative distinct values,
interpretable as being predicted by the many-worlds interpretation
of a cosmological quantum mechanics, each $\omega$-eigenvalue
describing the type of radiation that characterizes a different
parallel universe in a given multiverse scenario [8].

At this point, let us recapitulate. Variability of parameter
$\omega$ has been assumed as an intrinsic property af a generic
primordial universe in such a way that, while initially the size
of that universe increases, $\omega$ can take on a set of quantum
eigenvalues. The many-world interpretation of the resulting
quantum cosmological scenario is then adopted so that each
$\omega$-eigenstate would describe a differently expanding,
parallel universe characterized by a given $\omega$-dependent kind
of filling radiation. Finally, as the set of parallel primordial
universes enters the classical evolution regime, they continue to
keep the $\omega$-values fixed by the initial quantum dynamics,
and therefore all their characteristic properties remain settled
down forever, while noncausal connections among the classical
parallel universes are allowed to occur. Then, among the distinct
future destinations of the time travel of a primordial phantom
universe, we shall conjecture that anthropic principle [9] will
choose that time travelling which leads to a universe evolving
according to the flat Friedmann-Robertson-Walker dictum, filled
with radiation characterized by a positive parameter of the cosmic
equation of state $\omega=1/3$, which will be assumed to
correspond to one of the eigenstates of the
$\dot{\omega}$-representation obtained from solving the above
Wheeler-DeWitt equation. All those different destinations of the
time travel of the phantom universe which took place before the
big rip, or after it, for the same smaller than -1 equation of
state parameter (the same eigenstate) in the future of the same
phantom universe, or all of those corresponding to the future of
other negative or positive values of that parameter in different
parallel universes (other than the $\omega=1/3$ eigenstate), would
be aborted by the anthropic principle, relative to observers of
our present civilization.

We shall discuss next a rather simple classical model where it
will be considered how inflation can be implemented in an early
universe dominated by phantom energy according to the above lines.
Thus, whereas in the quantum cosmological initial era there is a
set of eigenstates (parallel universes), each characterized by a
particular eigenvalue of $\omega$, in the regime that follows that
initial era, we shall consider the classical evolution of each of
such parallel universes individually. We in this way interpret the
initial quantum regime as being characterized by the dynamic
quantum variables $\omega(t)$ and $a(t)$, and the subsequent
classical regime by a classical $a(t)$ and given constant values
of $\omega$. This is the way through which the quantum
cosmological evolution is linked to the classical scalar field
evolution. In what follows, we shall first regard an early
universe as being one of the above-mentioned parallel universes
filled with a dominating, generic dark energy fluid with constant
equation of state $p=\omega\rho$, and then particularize in the
special case where the universe is filled with phantom energy for
which $\omega<-1$, so ensuring a super-accelerated expansion
interpretable as an inflationary phase. Let us therefore take for
the Lagrangian of a dark energy field $\phi$ in a FRW universe the
customary general expression
\begin{equation}
L= \frac{1}{2}\dot{\phi}^2 -V(\phi),
\end{equation}
where $V(\phi)$ is the dark-energy field potential and we have
chosen the field $\phi$ to be defined in terms of a pressure $p$
and an energy density $\rho$ such that
\begin{equation}
\rho=\frac{1}{2}\dot{\phi}^2 +V(\phi) ,\;\;
p=\frac{1}{2}\dot{\phi}^2 -V(\phi) .
\end{equation}
So, for the equation of state $p=\omega\rho$ with constant
$\omega$, we have
\begin{equation}
\dot{\phi}^2=(1+\omega)\rho .
\end{equation}
Now, in our dark-energy dominated early universe, the Friedmann
equation for a flat geometry with scale factor $a(t)$ and constant
$\omega$ can be given by
\begin{equation}
\left(\frac{\dot{a}}{a}\right)^2 = \frac{8\pi G}{3}\rho ;
\end{equation}
thus, by integrating the equation for cosmic energy conservation,
$\dot{\rho}+3(p+\rho)\dot{a}/a=0$, and using Eq. (14), we can
obtain for the scale factor
\begin{equation}
a(t)=\left(C+\frac{3}{2}(1+\omega)t\right)^{2/[3(1+\omega)]}\equiv
T(t)^{2/[3(1+\omega)]} ,
\end{equation}
where $C\equiv C(\omega)$ is a constant for every value of
$\omega$. From Eqs. (13) and (15) we can then derive for the
potential of the dark energy field
\begin{equation}
V(\phi)=\frac{1}{2}\rho_0(1+\omega)e^{-3\sqrt{\frac{1+\omega}{\rho_0}}\phi}
,
\end{equation}
where $\rho_0$ is the initial arbitrary constant value of the
energy density, and the scalar field is given by
\begin{equation}
\phi=\frac{2}{3}\sqrt{\frac{\rho_0}{1+\omega}} \ln T(t) .
\end{equation}

We now specialize in the phantom region. For the phantom energy
regime $\omega<-1$ and $\phi\rightarrow i\Phi$ [10], Eqs. (16) and
(17) become
\begin{equation}
V(\Phi)=-\frac{1}{2}\rho_0(|\omega|-1)e^{3\sqrt{\frac{1+\omega}{\rho_0}}\Phi}
,
\end{equation}
\begin{equation}
\Phi=-\frac{2}{3}\sqrt{\frac{\rho_0}{|\omega|-1}} \ln T' ,
\end{equation}
with
\begin{equation}
T'=C(\omega<-1)-\frac{3}{2}(|\omega|-1)t .
\end{equation}

Using the phantom potential (18), the phantom field will initially
(at $t=0$) be at the bottom of the potential, being driven to
climb up alone the potential thereafter. That evolution of the
phantom field would take place while the universe rapidly inflates
according to the law $a=(T')^{-2/[3(|\omega|-1)]}$. The
inflationary process can only stop before reaching the big rip if
wormholes with a Planck-sized throats, originally placed in the
quantum spacetime foam, are allowed to exist and accrete phantom
energy. As a result of such an accretion process the wormhole
throats will increase at a rate quite greater than the one at
which the universe inflates, so that the wormholes eventually
become larger than the universe itself and finally reach an
infinite size at a finite time before the big rip given by [3]
\begin{equation}
\tilde{t}=\frac{t_*}{1+\frac{b_{0i}}{C(\omega<-1)\dot{b}_{0i}}} ,
\end{equation}
with $b_{0i}$ the initial wormhole throat radius, $t_*$ the
big-rip time,
\begin{equation}
t_*=\frac{2C(\omega<-1)}{3(|\omega|-1)} ,
\end{equation}
and
\begin{equation}
\dot{b}_{0i}=\frac{3}{4\pi^2 D\rho_0} ,
\end{equation}
in which $D$ is a constant of order unity. Thus, inflation of the
universe would stop due to non-chronal evolution once the causal
evolution reaches a time $\bar{t}$, before the big trip, at which
time the wormhole throat radius just overcomes the size of the
universe. For a constant equation of state parameter
$\omega=-5/3$, by instance, that time will be given by
\begin{equation}
\bar{t}=C(\omega=-5/3)-\left(\frac{A+\sqrt{A^2- \frac{4b_{0i}^2
C(\omega=-5/3)}{\dot{b}_{0i}t_*}}}{2b_{0i}}\right) ,
\end{equation}
where
\begin{equation}
A=1+\frac{b_{0i}}{\dot{b}_{0i}t_*} .
\end{equation}
After $t=\bar{t}$, the phantom universe as a whole enters a
noncausal phase where it can travel through time in a non-chronal
way, back to its origin or toward its future to get, relative to
present observers in our universe, into the observable Friedmann
cosmology we are familiar with. Among all the presumably infinite
number of possible future destinations of the primordial cosmic
time travel, which are allowed by the quantum parallel universe
picture, the anthropic principle would in this way choose only
that future evolution which is governed by the observable
Friedmann scenario dominated by radiation with $\omega=1/3$,
leaving the remaining potentially evolutionary cosmological
solutions as aborted possibilities, relative to present
human-being civilization (Fig. 2).

Since the general $\omega$-dependent expressions for radiation
temperature and energy density are respectively given by [11]
$\Theta\propto(1+\omega)a^{-3\omega}$ and
$\rho\propto[\Theta/(1+\omega)]^{(1+\omega)/\omega}$, and hence
the temperature of the phantom universe is negative and therefore
hotter than that of the host universe, time travelling from a
$\omega<-1$ universe to a $\omega=1/3$ universe (both assumed to
be different eigenstates of the same quantum-mechanical cosmic
wave equation, see Fig. 3) will naturally follow the natural flow
of energy and {\it globally} convert a radiation field which was
initially characterized by a negative temperature with very large
absolute value and a very large energy density, at $\omega<-1$,
into a radiation field at positive temperature with quite smaller
absolute value and quite smaller energy density, at $\omega=1/3$
(because cosmological effects dictated by the eigenvalue of
$\omega$ will prevail over microscopic, local effects globally).
At the same time that such a cosmological conversion takes place,
however, the phantom stuff should {\it locally} interact with the
stuff of the host Friedmann universe, as these stuffs should
initially preserve their microscopic and thermodynamic original
properties locally. In fact, within a given small volume $\delta
V$ filled with phantom energy the internal energy, $\delta
U=(1+\omega)\rho\delta V$, is definite negative and its absolute
value quite larger than that for the positive internal energy of
the radiation initially in the host universe, within the same
volume $\delta V$. As a result from that initial local
interaction, all of the energy initially present in the host
universe was annihilated, while almost all phantom radiation
remained practically unaffected microscopically, but would
globally behave like though if cosmologically it was the
$\omega=1/3$-radiation initially contained in the Friedmann
universe after having undergone an inflationary period, in spite
of the fact that a universe with $\omega=1/3$ by itself could
never undergo primordial inflation or reheating without including
an extra inflaton field, which we do not assume to exist. Thus,
for a current observer, the transferred phantom radiation left
after microscopic interaction would just be the customary
microwave background radiation characterized by a parameter
$\omega=1/3$.

It follows that, relative to current observers, besides
transferring a subdominant proportion of matter, the overall neat
observable effect induced on the Friedmann universe when a whole
phantom universe is transferred by a time travel act into it would
in practice be converting the relative-to-current-observers causal
disconnectedness owing to the inflationless universe into full
causal connectedness of all of its components, relative to such
observers, so solving any uniformity and horizon problems (Fig.
4). In this way, we would no longer expect ourselves to be made
just of the matter and positive energy created at our own
universe's big bang, but rather out of a mixture of such
components with the matter and phantom energy created at the
origin of a universe other than ours. It is worth remarking that
for the considered time travel to take place it must be performed
before the wormhole throat radius becomes infinity (that is along
the time interval $\Delta_{nonchronal}$ in Fig. 1), since
otherwise the wormhole is converted into a Einstein-Rosen bridge
whose throat would immediately pinches off, leaving a pair
black-white holes which accrete phantom energy and rapidly vanish
[3] (see Fig. 5).

We shall finally notice that, relative to a "quantum" observer
(i.e. an observer able to simultaneously observe all the
eigenuniverses), the whole process we have just discussed violates
the second law of thermodynamics. In fact, if we disregard the
matter contents of the universes, then the entropy of each of
these universes is a given universal constant $\sigma$ [11], and
therefore the total initial entropy would be $n\sigma$, with $n$
the number of eigenuniverses. After the above discussed time
travel, the total entropy would decrease down to a value
$(n-1)\sigma$. Moreover, a violation of the second law would also
take place relative to an observer in our universe due to the
increase of coherence induced in the host universe by exchanging
its original radiation content for that carried into it by the
phantom universe. These violations of the second law can be
thought to be the consequence from the fact that any stuff having
phantom energy must be regarded as an essentially quantum stuff
with no classical analog, where negative temperatures and
entropies are commonplace, such as has been seen to occur in
entangled-state and nuclear spin systems.

The whole scenario discussed in this paper is quite speculative
and will of course pose some problems and difficulties which would
deserve thorough consideration. The most important problem refers
to density and gravitational wave fluctuations which are
originated in the phantom inflationary epoch, and the way these
predicted fluctuations compare with the temperature fluctuations
currently observed in WMAP CMB [12]. Though we do not intend to
perform a complete investigation on such subjects in this paper, a
brief discussion appears worth considering. From Eqs. (15), (18) -
(20), it can be shown that the slow-climb conditions for our
phantom inflation model are
\begin{equation}
\epsilon_{{\rm ph}}=-\frac{\dot{H}}{H^2}=\frac{3}{2}(|\omega|-1)
<< 1 ,\;\;\; \delta{{\rm ph}}=-\frac{\ddot{\Phi}}{\dot{\Phi}H}=
\epsilon_{{\rm ph}} << 1 ,
\end{equation}
where $H=\dot{a}/a$. When conditions (26) are satisfied the scale
factor will approximately evolve with time in an exponential, de
Sitter like way, while the parameter $\omega<-1$ must take on
values which are very close to -1. Metric perturbations can then
be studied following e.g. the procedure in Ref. [2]. Thus, in the
longitudinal gauge and absence of anisotropic stresses, the scalar
metric perturbations can be expressed in terms of a perturbed
metric as a function of both the Bardeen potential, $V_B$, and the
conformal time $\eta=\int dt/a(t)$, with
$a(\eta)=\left[-(|\omega|-1)\eta/2\right]^{-2/[2(|\omega|-1)]}$.
As expressed in terms of the Bardeen potential as well, the
curvature perturbation on uniform comoving hypersurface is then
given by
\begin{equation}
\varsigma =V_B +\frac{(3|\omega|-1)\eta V_B '-2V_B}{3(|\omega|-1)}
,
\end{equation}
where $'=d/d\eta$. It follows that the differential equation of
motion for the perturbation $\varsigma$ can be written as
\begin{equation}
\vartheta_k '' +\left(k^2 -\frac{2(1+3|\omega|)}{(3|\omega|-1)^2
\eta^2}\right)\vartheta_k = 0,
\end{equation}
where $\vartheta=a^2\Phi '\varsigma/a'=z\varsigma$, with
$z=\sqrt{\rho_0(|\omega|-1)}\left(-(3|\omega|-
1)\eta/2\right)^{-2/(3|\omega|-1)}$. Eq. (28) admits an exact
analytical solution in terms of the Bessel function $B$ [13]:
\begin{equation}
\vartheta_k =\eta^{1/2}B_{\frac{3(1+|\omega|)}{2(3|\omega|-
1)}}\left(k\eta\right)\simeq \eta^{1/2}B_{\frac{3}{2}+
\epsilon_{{\rm ph}}}\left(k\eta\right),
\end{equation}
the last approximate equation being valid if the slow-climb
conditions are satisfied. The inclusion of suitable initial
conditions will choose the precise Bessel function we should use.
The amplitude associated with this perturbation can then be given
by [2]
\begin{equation}
A_s =\frac{2}{3(|\omega|-1)(T')^2_{k=aH}} ,
\end{equation}
while the spectral index becomes
$n_s=\frac{3}{2}\left(\frac{5|\omega|-3}{3|\omega|-1}\right)$. The
amplitude for tensor perturbation can be analogously derived. It
reads
\begin{equation}
A_t=\left(\frac{1}{T'}\right)^2_{k=aH} ,
\end{equation}
with a spectral index $n_t\simeq-3(|\omega|-1)$.

The ratio $A_t/A_s=3(|\omega|-1)/2\simeq\epsilon_{{\rm ph}}$
should now be compared with the tensor/scalar ratio of CMB
quadrupole contributions, $r\sim\epsilon_{{\rm ph}}$, to low
order. Our phantom scenario with $\epsilon_{{\rm ph}}=\delta_{{\rm
ph}}$ and exponential potential differs from other phantom
inflation models [2], and a thorough join comparation of all of
these models with the usual scalar field inflationary scenarios on
the $r-n_s$ plane is left for a future publication. Actually, just
like it happens with scalar field models for inflation, the
phantom inflation scenarios could only be taken seriously if they
succeeded in predicting the types of temperature fluctuations
discovered from the WMAP CMB data.

\vspace{.8cm}

\noindent{\bf Acknowledgements} The author thanks Carmen L.
Sig\"{u}enza for useful information exchange and comments. This work
was supported by DGICYT under Research Project No. BMF2002-03758.

\pagebreak

\noindent\section*{References}

\begin{description}

\item [1] J.S. Alcaniz, Phys. Rev. D69, 083521 (2004) , and
references therein.

\item [2] Y-S. Piao and Y.-Z. Zhang, astro-ph/0401231

\item [3] P.F. Gonz\'{a}lez-D\'{\i}az, astro-ph/0404045, Phys. Rev.
Lett. (in press, 2004)

\item [4] Visser, {\it Lorentzian Wormholes} (AIP Press,
WoodburyNew York, USA, 1995); S. Nojiri, O. Obregon, S.D. Odintsov
and K.E. Osetrin, Phys. Lett. B449, 173 (1999).

\item [5] P.F. Gonz\'{a}lez-D\'{\i}az, Phys. Rev. D68, 084016
(2003).

\item [6] R.R. Caldwell, Phys. Lett. B545, 23 (2002); R.R.
Caldwell, M. Kamionkowski and N.N. Weinberg, Phys. Rev. Lett. 91,
071301 (2003); P.F. Gonz\'{a}lez-D\'{\i}az, Phys. Rev. D68, 021303 (2003);
J.D. Barrow, Class. Quant. Grav. 21, L79 (2004); M. Bouhmadi and
J.A. Jim\'{e}nez- Madrid, astro-ph/0404540 .

\item [7] J.B. Hartle and S.W. Hawking, Phys. Rev. D28, 2960
(1983); S.W. Hawking, Phys. Rev. D32, 2489 (1985); S.W. Hawking,
Quantum Cosmology, in: {\it Relativity Groups and Topology}, Les
Houches Lectures, edited by B. DeWitt and R. Stora (North-Holland,
1984).

\item [8] See the contributions in the book {\it The Many-Worlds
Interpretation of Quantum Mechanics}, edited by B.SDeWitt and N.
Graham (Princeton University Press, Princeton, New Jersey, USA,
1973).

\item [9] J.D. Barrow and F.J. Tipler, {\it The Anthropic
Cosmological Principle} (Clarendon Press, Oxford, UK, 1986).

\item [10] P.F. Gonz\'{a}lez-D\'{\i}az, Phys. Rev. D69, 063522
(2004).

\item [11] P.F. Gonz\'{a}lez-D\'{\i}az and C.L. Sig\"{u}enza, Phys. Lett.
B589, 78 (2004).

\item [12] W.K. Kinney, E.W. Kolb, A. Melchiorri and A. Riotto,
Phys. Rev. D69, 103516 (2004)

\item [13] M. Abramowitz and I.A. Stegun, {\it Handbook of
Mathematical Functions} (Dover, New York, USA, 1965).

\end{description}

\begin{figure}
\includegraphics[width=.9\columnwidth]{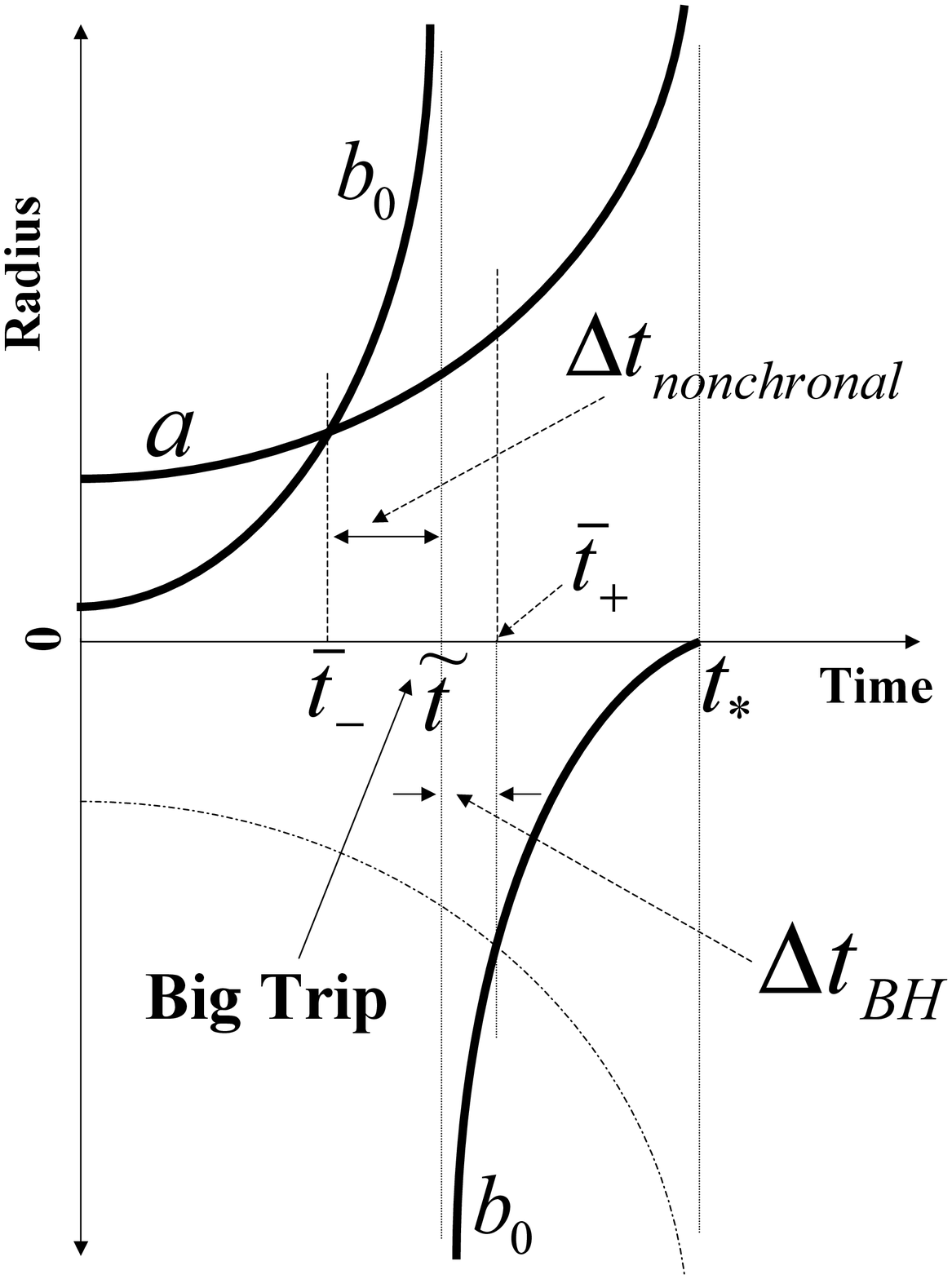}
\caption{\label{fig:epsart} Evolution of the radius of the
wormhole throat, $b_0$, induced by accretion of phantom energy. At
time $t=\tilde{t}$, the negative exotic mass becomes infinite and
then changes sign, so converting the wormhole into an
Einstein-Rosen bridge whose associated mass decreases down to zero
at the big rip at $t=t_*$. During the time interval $\Delta
t_{nonchronal}$ there will be a disruption of the causal evolution
of the whole universe. The evolution of the phantom universe at
time $t=\tilde{t}$ is fully noncausal and all sorts of time travel
are then allowed. Therefore the moment at $t=\tilde{t}$ is here
dubbed as the "Big Trip".}
\end{figure}

\begin{figure}
\includegraphics[width=.9\columnwidth]{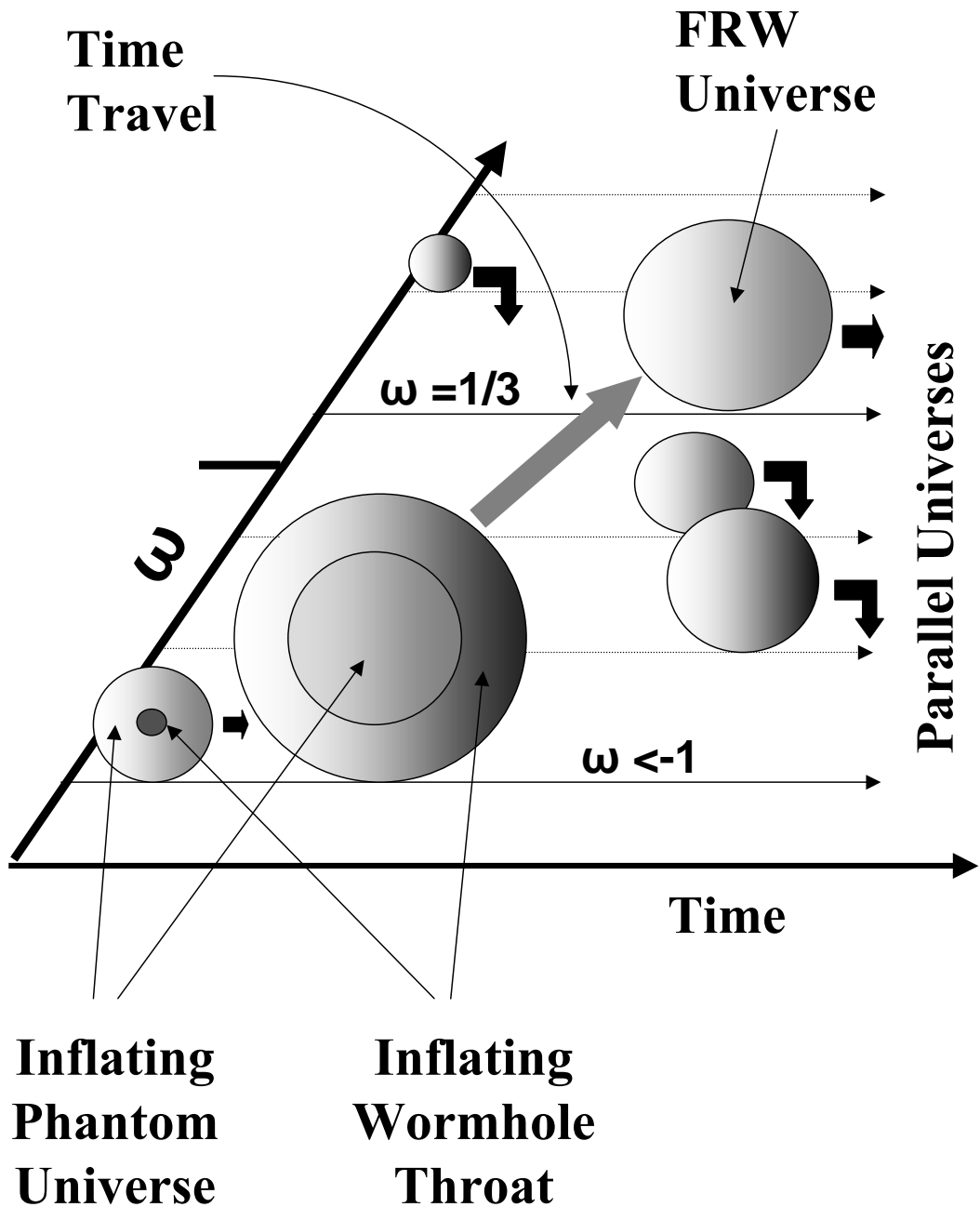}
\caption{\label{fig:epsart} Pictorial representation of the smooth
exit from the primordial inflationary process driven by phantom
energy. It is based on the time-travelling of a whole,
sufficiently inflated, universe from the phantom-energy dominated
regime to the usual radiation-dominated regime. Such a nonchronal
process is chosen by anthropic arguments over all other possible
cosmological final configurations allowed by the many-worlds
interpretation of the quantum mechanics of the early universe.}
\end{figure}

\begin{figure}
\includegraphics[width=.9\columnwidth]{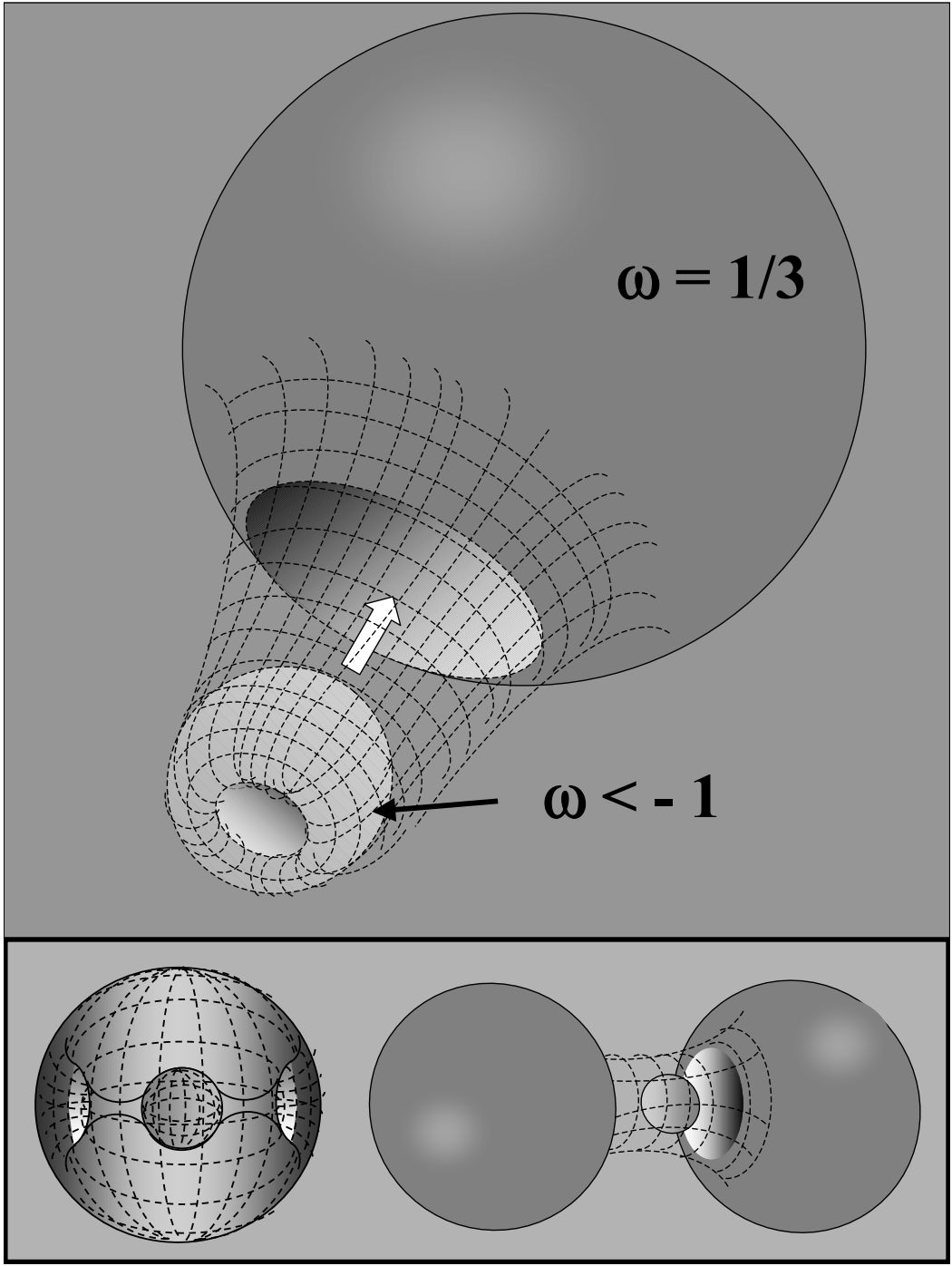}
\caption{\label{fig:epsart} Pictorial representation of the time
travel process by which a phantom universe with $\omega<-1$ gets
in a universe with $\omega=1/3$ in its future. That process is
carried out by using the topology represented in upper part of
this figure by which an inflated wormhole whose size has exceeded
the radius of the phantom universe itself inserts one of its
mouths in the phantom universe while its other mouth is inserted
in the host universe. In the inset at the bottom of the figure we
also show the other two possible topologies that the larger
wormhole throat and the phantom universe may adopt. On the left it
is the topology involving just the phantom universe and the
wormhole, and on the right we show the most complicated topology
where, besides the phantom universe and the wormhole, there are
two extra universes in which the two wormhole mouths are
respectively inserted.}
\end{figure}

\begin{figure}
\includegraphics[width=.9\columnwidth]{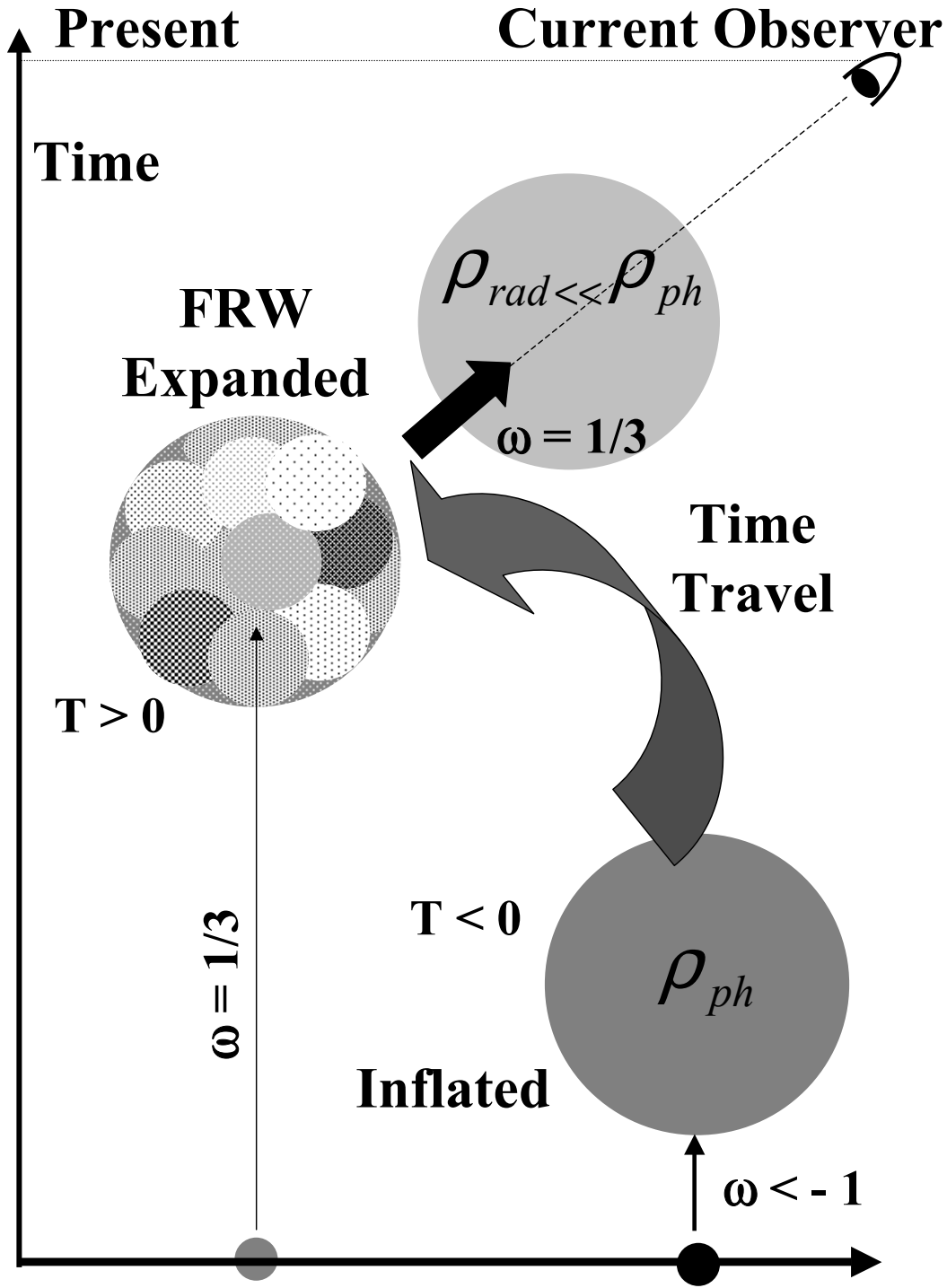}
\caption{\label{fig:epsart} Uniformity is transferred from the
inflated phantom universe to our universe when the former universe
is absorbed by the latter universe after the time travel. Before
the moment of time travelling, our universe was completely
causally disconnected with respect to current observers. As the
phantom universe was hotter than ours, when they came in contact
all the uniformly distributed energy in it was transferred into
our universe, annihilating all its inhomogeneities and thereby
disappearing.}
\end{figure}

\begin{figure}
\includegraphics[width=.9\columnwidth]{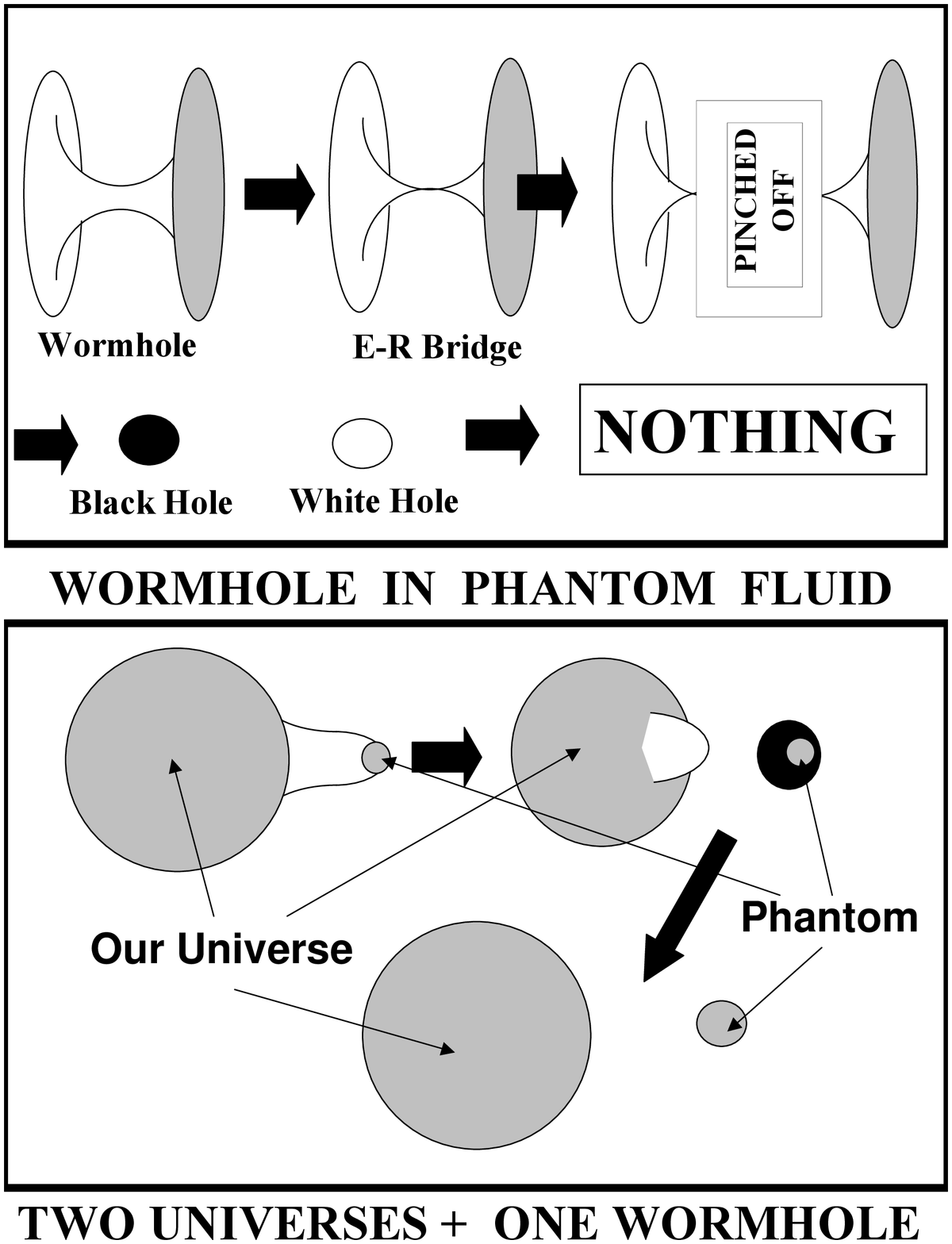}
\caption{\label{fig:epsart} Upper part: After time $t=\tilde{t}$ a
single wormhole is immediately converted into a Einstein-Rosen
bridge whose throat rapidly pinches off, leaving a black hole plus
a white hole. These holes will then fade rapidly off by continuing
accreting phantom energy. Lower part: The same regime for the case
of a wormhole connecting a phantom universe with another distinct
universe like ours. In this case, each of these two universes is
first enclosed in a giant black hole or white hole which are
mutually disconnected. Then, the holes fade rapidly off and
finally the two disconnected universes are left as the sole result
of the whole process.}
\end{figure}

\end{document}